\documentclass[conference]{IEEEtran}
\usepackage{amsfonts}
\usepackage{amssymb}
\usepackage{cite}
\usepackage[cmex10]{amsmath}
\usepackage{float}
\usepackage{color}
\usepackage{stfloats,fancyhdr}

\usepackage{algorithm}
\usepackage{algorithmic}
\usepackage{multirow}
\usepackage{changepage}
\usepackage[normalem]{ulem}

\newtheorem{remark}{Remark}

\IEEEoverridecommandlockouts

\ifCLASSINFOpdf
  \usepackage[pdftex]{graphicx}
  \DeclareGraphicsExtensions{.pdf,.jpeg,.png}
\else
  \usepackage[dvips]{graphicx}
  \DeclareGraphicsExtensions{.eps}
\fi

\usepackage{subfigure}

\usepackage{fancybox,dashbox}

\begin{document}

\title{Wireless-Powered Two-Way Relaying with Power Splitting-based Energy Accumulation
\thanks{This research was supported by ARC grants DP150104019 and
FT120100487. The research was also supported by funding from the Faculty
of Engineering and Information Technologies, The University of Sydney,
under the Faculty Research Cluster Program and the Faculty Early Career
Researcher Scheme.}
}

\author{Yifan Gu, He Chen, Yonghui Li, and Branka Vucetic
\\
School of Electrical and Information Engineering, The University of Sydney, Sydney, NSW 2006, Australia\\
Email: \{yifan.gu, he.chen, yonghui.li, branka.vucetic\}@sydney.edu.au
}



\maketitle

\begin{abstract}
This paper investigates a wireless-powered two-way relay network (WP-TWRN), in which two sources exchange information with the aid of one amplify-and-forward (AF) relay. Contrary to the conventional two-way relay networks, we consider the scenario that the AF relay has no embedded energy supply, and it is equipped with an energy harvesting unit and rechargeable battery. As such, it can accumulate the energy harvested from both sources' signals before helping forwarding their information. In this paper, we develop a power splitting-based energy accumulation (PS-EA) scheme for the considered WP-TWRN. To determine whether the relay has accumulated sufficient energy, we set a predefined energy threshold for the relay. When the accumulated energy reaches the threshold, relay splits the received signal power into two parts, one for energy harvesting and the other for information forwarding. If the stored energy at the relay is below the threshold, all the received signal power will be accumulated at the relay's battery. By modeling the finite-capacity battery of relay as a finite-state Markov Chain (MC), we derive a closed-form expression for the system throughput of the proposed PS-EA scheme over Nakagami-m fading channels. Numerical results validate our theoretical analysis and show that the proposed PS-EA scheme outperforms the conventional time switching-based energy accumulation (TS-EA) scheme and the existing power splitting schemes without energy accumulation.
\end{abstract}


\IEEEpeerreviewmaketitle

\section{Introduction}
Radio-frequency (RF) energy transfer and harvesting technique has been proposed to prolong the lifetime of energy constrained networks recently. It enables wireless devices to extract the energy carried by RF signals broadcast by ambient/dedicated energy transmitters to charge their batteries. This cutting-edge technology has opened a new research paradigm, termed wireless-powered communication (WPC), WPC has been intensively studied in conventional point to point or point to multiple point networks. Recently WPC has been also extended to relay networks \cite{powersplitting3,CHEN_TWC}. In \cite{powersplitting3}, Nasir \emph{et al.} first studied the idea of WPC in a three-node relay network, in which a wireless-powered relay uses the energy harvested from source to forward its information. In \cite{powersplitting3}, two practical relaying protocols, namely time switching (TS) relaying and power splitting (PS) relaying were proposed and analyzed. Specifically, TS relaying splits the received signal in time domain by a switcher while PS relaying splits the signal in power domain by a power splitter. After that one signal is fed to energy receiver for energy harvesting and the other one is delivered to information receiver for information forwarding. \cite{CHEN_TWC} extended \cite{powersplitting3} into a more general scenario with multiple interfering WPC relay links.

Two-way relay network (TWRN) has attracted a lot of research interests over the past decade \cite{Raymond-two-way-relay}. Compared with one-way relaying, two-way relaying can enhance the network throughput and spectral efficiency. Motivated by this fact, the design and analysis of wireless-powered two-way relay networks (WP-TWRNs) have attracted considerable research interests recently \cite{TW-TS-1,TW-TS-2,TW-TS-3,TW-PS-1,TW-PS-2,TW-PS-3,TW-PS-4}. However, existing works on WP-TWRNs assumed that the wireless-powered relay exhausts the harvested energy in each transmission block to perform information transmission/forwarding. This assumption may lead to only a suboptimal system performance due to the following facts. Firstly, when the receiving channels of the wireless-powered relay suffer from deep fading, it can only harvest a little amount of energy. Due to the symmetric characteristic of two-way relay channels, the channels from the relay to the sources are also in poor condition and thus the relay may not be able to perform an effective information transmission/forwarding even it exhausts all the harvested energy. On the other hand, when the receiving channels of the two-way relay are in good condition, it should use only part of its harvested energy for information transmission/forwarding and save the rest energy for future use. In this sense, the consideration and modeling of the energy accumulation (EA) process is essential such that the wireless-powered relay is able to accumulate the harvested energy and perform information transmission in an appropriate time with efficient power levels.

Motivated by the aforementioned problem, in this paper, we focus on the design and analysis of a WP-TWRN with EA. Specifically, we consider a WP-TWRN consisting of two sources and one wireless-powered relay. By noting that the PS technique considerably has better performance than the TS technique \cite{powersplitting3}, we assume that the PS technique is used at the wireless-powered relay. To the best knowledge of the authors, this is the first work that design and analyze a WP-TWRN with PS and EA.

It should be noted that the Markov Chain (MC)-based method has been used to characterize the performance of wireless-powered communications with EA for point to point networks, see \cite{row-stac,TS-accumulation,five-multi-relay,YIFAN11}. In all these works, the TS-based EA was considered such that the battery of wireless-powered node(s) can only be either charged or discharged during each transmission block. However, for the considered system with PS-based EA, the relay may experience both energy harvesting and information forwarding operations during each transmission block by splitting the received signal power. This will lead to both charging and discharging of the battery and it is actually hard to determine whether the relay battery is finally charged or discharged during the block. In this sense, the existing MC-based analysis methods are no longer applicable and a new analytical approach should be proposed to characterize the system performance of the considered WP-TWRN.

The main contributions of this paper are summarized as follows: {\textbf{(1)}} We develop a power splitting-based energy accumulation (PS-EA) scheme for the considered WP-TWRNs. {\textbf{(2)}} We model the dynamic charging/discharging behaviors of the relay battery by a finite-state Markov chain (MC). We then propose a novel mode-based method to evaluate the transition matrix and stationary distribution of the MC. {\textbf{(3)}} Considering an amplify-and-forward protocol at relay, we derive a closed-form expression for the system throughput of the proposed PS-EA scheme over Nakagami-m fading channels. This is in contrast to the ideal Rayleigh fading model used in the existing works \cite{TW-TS-1,TW-TS-2,TW-TS-3,TW-PS-1,TW-PS-2,TW-PS-3,TW-PS-4,row-stac,TS-accumulation,five-multi-relay}. The system throughput is defined as the sum of average throughput of the two sources. All theoretical analysis is then validated by numerical results. It is shown that the proposed PS-EA scheme can considerably outperform its counterpart time switching-based energy accumulation (TS-EA) scheme and the traditional PS scheme without EA.

\textbf{\emph{Notation}}: Throughout this paper, we use $f_{X}(x)$ and $F_{X}(x)$ to denote the probability density function (PDF) and cumulative distribution function (CDF) of a random variable $X$. $\Gamma \left( {\cdot} \right)$ is the Gamma function \cite[Eq. (8.310)]{Tableofintegral}, $\gamma \left( {\cdot,\cdot} \right)$ is the lower incomplete gamma function \cite[Eq. (8.350.1)]{Tableofintegral} and $K_v \left( {\cdot} \right)$ is the modified bessel function of the second kind \cite[Eq. (8.407)]{Tableofintegral}. $\left( \cdot \right)^T$ represents the transpose of a matrix or vector and ${\bf I}$ denotes the identity matrix.

\section{System Model and Protocol Description}
In this paper, we consider a WP-TWRN consisting of two sources and one amplify-and-forward (AF) relay, which is deployed to assist the information exchange between two sources. We assume that there is no direct link between two sources due to obstacles or severe attenuation. Also, all nodes are equipped with single antenna and work in half-duplex mode. Unlike conventional networks where the relay has embedded power supply, we consider the scenario that the relay is a wireless-powered device and purely rely on the energy harvested from RF signals broadcast by two sources to perform information forwarding. Moreover, the relay is equipped with a finite-capacity rechargeable battery such that it can perform energy accumulation and scheduling across different transmission blocks.

We use subscript-$A$, $B$ and $R$ to denote two sources and relay, respectively. We note that the up-to-date wireless energy harvesting techniques could only be operated within a relatively short communication range such that the line-of-sight (LoS) path is very likely to exist in these links. In this sense, the Rician fading model should be used to characterize the channel fading for links between two sources and relay. However, the statistical functions (e.g., CDF and PDF) of Rician fading are very complicated, which would make the the performance analysis of the proposed scheme extremely difficult \cite{Zhong_Tcom_2015_Wireless}. Fortunately, the Rician distribution could be well approximated by the more tractable Nakagami-m fading model. Thus, in this paper we assume that links $A-R$ and $B-R$ are subject to Nakagami-m fading with fading severity parameters\footnote{For the purpose of exploration, we consider the case that $m_A$ and $m_B$ are integers but their values can be different.} $m_A$, $m_B$ and average power gains ${\Omega _A}$, ${\Omega _B}$ respectively. Besides, both channels are assumed to experience slow, independent, and frequency-flat fading such that instantaneous channel gains remain unchanged within each transmission block but change independently from one block to the other.

Subsequently, we design a power splitting-based energy accumulation (PS-EA) scheme for the considered WP-TWRN. The received signals at the relay from both sources can be split into two parts by a power splitter: one for energy harvesting and the other for information forwarding. Let $T$ denote the duration of each transmission block, which is further divided into two time slots with equal length $T/2$. At the beginning of each transmission block, the relay chooses to operate in one of the two possible modes (denoted by Mode I and Mode II) based only on its current battery status. To determine whether the relay has accumulated sufficient energy, we set a predefined energy threshold $P_R T/2$ for the relay, which can support a forwarding transmit power of $P_R$ in the second time slot. Mode I operation is invoked when the stored energy is below the energy threshold. Otherwise, Mode II operation is activated to forward the received information. When the relay operates in Mode I, during the first time slot, the received signal from $A$ and $B$ is used entirely for energy harvesting to charge the relay battery. For operation simplicity, we restrict that the relay provides no feedback to the sources that the sources always remain in silence during the second time slot. Thus, the relay choosing Mode I operation will not harvest energy from the sources during the second time slot. In Mode II, the received signal during the first time slot is split into two parts, one for energy harvesting and the other for information forwarding. In the second time slot, the relay will amplify and forward the received signal to two sources with a transmit power of $P_R$.

Let $P_A$, $P_B$ denote the transmit power of $A$, $B$ and $x_A$, $x_B$ denote the transmitted symbol of $A$ and $B$ with unit energy respectively. The received signal at the relay during the first time slot is given by
\begin{equation}
y_{R} = \sqrt{P_A} h x_A + \sqrt{P_B} g x_B+n_{1},
\end{equation}
where $h$ is the channel coefficient between $A$ and $R$, $g$ is the channel coefficient between $B$ and $R$ and $n_{1} \sim \mathcal{CN}\left( {0, N_1} \right)$ denotes the narrow-band Gaussian noise introduced by the receiving antenna \cite{powersplitting3}.

Without loss of generality, we consider a normalized transmission block (i.e., $T=1$) hereafter. When the relay works in Mode I, all the received power is used for energy harvesting and the amount of harvested energy at $R$ during the first time slot can be expressed as
\begin{equation}\label{EH1}
{\tilde E_{\rm I}} =  {1 \over 2}\eta \left({P_A}{H}+{P_B}{G}\right),
\end{equation}
where $0 < \eta  < 1$ is the energy conversion efficiency and ${H}=  \left| {h } \right|^2$, ${G}=  \left| {g} \right|^2$ is the channel power gain of the considered links. Note that in (\ref{EH1}), we ignore the amount of energy harvested from the noise since the noise power is normally very small and below the sensitivity of the energy receiver. As the relay will not forward information in Mode I, the output signal-to-noise ratios (SNRs) at sources $A$ and $B$ are $\gamma_A=\gamma_B =0$.

On the other hand, if $R$ opts to operate in Mode II, the harvested energy is given by
\begin{equation}\label{EH2}
{\tilde E_{\rm II}} =  \lambda{\tilde E_{\rm I}}={1 \over 2}\lambda \eta \left({P_A}{H}+{P_B}{G}\right),
\end{equation}
where $0<\lambda<1$ is the power splitting coefficient for energy harvesting. The broadcast signal by the relay node can be written as
\begin{equation}
x_{R} = \Lambda \sqrt{P_R} \left( \sqrt{1-\lambda}y_{R}+n_2\right),
\end{equation}
where $n_{2} \sim \mathcal{CN}\left( {0, N_2} \right)$ is the AWGN introduced by the signal conversion from passband to baseband \cite{powersplitting3} and the power constraint factor is given by
\begin{equation}
\Lambda {\rm{ = }}{{\rm{1}} \over {\sqrt {\left( {{\rm{1 - }}\lambda } \right)\left( {{P_A}H{\rm{ + }}{P_B}G + {N_1}} \right) + {N_2}} }}.
\end{equation}
At the end of the Mode II, the received signals at source $A$ and $B$ can be expressed as
\begin{equation}
y_{A} = h x_{R} + n_A,
\end{equation}
\begin{equation}
y_{B} =g x_R + n_B,
\end{equation}
where $n_{A}, n_{B} \sim \mathcal{CN}\left( {0, N_0} \right)$ is the AWGN noise at the sources. Since each source node has perfect knowledge of its transmitted symbol, it then cancels the self interference term. After some mathematical manipulation, we can obtain the received SNRs at source $A$ and $B$ given by
\begin{equation}\label{SNRA}
\gamma_A = {{{{\overline \gamma  }_R}{{\overline \gamma  }_B}GH} \over {\left( {{{\overline \gamma  }_R} + {{\overline \gamma  }_A}} \right)H + {{\overline \gamma  }_B}G + 1}},
\end{equation}
\begin{equation}\label{SNRB}
\gamma_B = {{{{\overline \gamma  }_R}{{\overline \gamma  }_A}HG} \over {\left( {{{\overline \gamma  }_R} + {{\overline \gamma  }_B}} \right)G + {{\overline \gamma  }_A}H + 1}},
\end{equation}
where ${\overline \gamma  _A} = {{\left( {1 - \lambda } \right){P_A}} \over {\left( {1 - \lambda } \right){N_1} + {N_2}}}$, ${\overline \gamma  _B} = {{\left( {1 - \lambda } \right){P_B}} \over {\left( {1 - \lambda } \right){N_1} + {N_2}}}$ and ${\overline \gamma  _R} = {{{P_R}} \over {{N_0}}}$.

\section{Performance Analysis}
To analyze the performance of the proposed PS-EA scheme, in this section we characterize the dynamic behaviors of the relay battery. We follow \cite{row-stac} and consider a discrete-level and finite-capacity battery model. Thus, it is natural to use a finite-state Markov chain (MC) to model the dynamic behaviors of relay battery. It should be noted that the existing MC-based analysis methods are no longer applicable in our case due to the complicated charging and discharging behavior of the considered PS technique. A new mode-based approach is proposed to characterize the transition matrix and stationary distribution of the MC. Based on this, a closed-form expression of the system throughput of the proposed PS-EA scheme is derived.
\subsection{Markov Model of Relay Battery}

Let $C$ denote the capacity of the relay battery and $L$ denote the number of discrete energy levels excluding the empty level. Then, the $i$-th energy level of relay battery can be expressed as ${\varepsilon_i} = iC/L$, $i  \in \left\{  0,1,2 \cdots L\right\}$. As shown in \cite{Huang_TWC_2008_Wire}, the adopted discrete battery model can tightly approximate its continuous counterpart when the number of energy levels (i.e., $L$) is sufficiently large. We define state ${S_i}$ as the relay residual energy in the battery being ${\varepsilon_i}$. The transition probability $T_{i,j}$ is defined as the probability of transition from state $S_i$ to state $S_j$. With the adopted discrete-level battery model, the amount of harvested energy can only be one of the discrete energy levels. The discretized amount of harvested energy at the relay during Mode I and II is defined as
\begin{equation}\label{deh}
{E_{\rm K}} \buildrel \Delta \over = {\varepsilon _{j}}, \quad {j} = \arg \mathop {\max }\limits_{i \in \left\{ {0,1, \cdots ,L} \right\}} \bigg\{{\varepsilon _i}:{\varepsilon _i} \le {\tilde E_{\rm K}}\bigg\},
\end{equation}
where $\rm K \in \left\{\rm I, \rm II\right\}$. On the other hand, in Mode II, we define the discretized transmitted energy for the relay as one of $L$ energy levels of the battery excluding the empty level $P_R /2 \in \left\{ \varepsilon_1, \varepsilon_2, \cdots, \varepsilon_L \right\}$. For convenience, let integer $\delta = {P_R \over {2\varepsilon_1}} \in \left\{ 1, 2, \cdots, L \right\}$ denote the energy level of the relay transmit power $P_R$.


\subsection{Transition Matrix}
We now evaluate the state transition probabilities of the MC at relay. We notice that the transition behaviors of the formulated MC for relay battery actually depend on the operation modes of the proposed PS-EA scheme. The relay battery is charged during each Mode I operation and it is first charged then discharged during each Mode II operation. Motivated by this, we propose a novel mode-based approach to evaluate the transition probabilities of the MC of relay battery. In the proposed approach, we summarize the calculations of all possible transition probability into the following two cases.
\subsubsection{The relay operates in Mode \rm I (${S_i}\quad to\quad {S_j}$ with $0 \le i  < \delta $ and $\forall j$)}\label{an1}
When the relay operates in Mode I, it harvests energy from the two sources and transits from state $S_i$ to $S_j$, $j \in \left\{i,i+1,\cdots, L\right\}$ due to the fact that the battery is not discharged. Specifically, $j=i$ represents the case where the harvested energy is discretized to zero and the battery remains the same, and $j=L$ denotes the case that the battery is fully charged by two sources during the first time slot. From the definition of discretization given in (\ref{deh}), the transition probability is given by
\begin{equation}
\begin{split}
&{T_{i,j}}= \left\{ {
\begin{matrix}
\begin{split}
   &{\Pr \left\{ {{E_{\rm I}}} = {\varepsilon _{j-i}} \right\}, \quad \text{if} \quad i\le j<L}\\
  &{\Pr \left\{ {{E_{\rm I}}} > {\varepsilon _{L-i}} \right\}, \quad \text{if} \quad j=L}\\
  &{0, \quad \text{Otherwise}}\\
\end{split}
\end{matrix}
  } \right.\\
  & \quad =\left\{ {
\begin{matrix}
\begin{split}
   &{\Pr \left\{  {\varepsilon _{j-i}}\le {{\tilde E_{\rm I}}} < {\varepsilon _{j-i+1}} \right\} , \quad \text{if} \quad i\le j<L}\\
  & {\Pr \left\{ {{\tilde E_{\rm I}}} > {\varepsilon _{L-i}} \right\}, \quad \text{if} \quad j=L}\\
  &{0, \quad \text{Otherwise}}\\
\end{split}
\end{matrix}
  } \right..\\
\end{split}
\end{equation}

To proceed, we need to characterize the distribution of the harvested energy in Mode I given in (\ref{EH1}), which is the sum of two gamma random variables with parameters $m_{A}$, $m_{B}$ and average power gains $\Omega_1 = {1 \over 2}\eta{P_A}{\Omega _{A}}$, $\Omega_2 = {1 \over 2}\eta{P_B}{\Omega _{B}}$ \cite[(Eq.2.21)]{fadingchannels}. The CDF of the sum of gamma random variables has been widely studied in the existing literature. For presentation brevity, we omit the expression of ${F_{\tilde E_{\rm I}}}\left(x\right)$ and it can be found in \cite[Eq. (9)]{sum-two-gamma}.
The transition probability for this case can now be summarized as
\begin{equation}
\begin{split}
&{T_{i,j}} =\left\{ {
\begin{matrix}
\begin{split}
   &{{F_{\tilde E_{\rm I}}}\left( \varepsilon_{j-i+1} \right)-{F_{\tilde E_{\rm I}}}\left( \varepsilon_{j-i} \right) , \quad \text{if} \quad i\le j<L}\\
  & {1- {F_{\tilde E_{\rm I}}}\left( \varepsilon_{L-i} \right), \quad \text{if} \quad j=L}\\
  &{0, \quad \text{Otherwise}}\\
\end{split}
\end{matrix}
  } \right..\\
\end{split}
\end{equation}

\subsubsection{The relay operates in Mode \rm II (${S_i}\quad to\quad {S_j}$ with $ \delta \le i \le L $ and $\forall j$)}In this transition case, the relay first harvests energy from two sources through power splitting technique, then forwards the signal to the sources by consuming $\delta$ energy levels of the battery. Since after the energy harvesting phase, the energy level of the battery could be varied from level $i$ to level $L$ based on different amount of harvested energy as in Mode I. After the discharging phase, the end state of this case should fall into the set $S_j$, $j \in \left\{i-\delta,i+1-\delta,\cdots, L-\delta \right\}$. With the CDF of ${\tilde E_{\rm I}}$, the transition probability of this case can now be calculated and expressed in a closed-form shown in (\ref{transition}) on top of next page.
\begin{figure*}[!t]
\begin{equation}\label{transition}
\begin{split}
&{T_{i,j}}  =\left\{ {
\begin{matrix}
\begin{split}
   &{\Pr \left\{ {{E_{\rm II}}} = {\varepsilon _{j-i+\delta}} \right\}, \quad \text{if} \quad i- \delta \le j<L-\delta}\\
  &{\Pr \left\{ {{E_{\rm II}}} > {\varepsilon _{L-i}} \right\}, \quad \text{if} \quad j=L-\delta}\\
  &{0, \quad \text{Otherwise}}\\
\end{split}
\end{matrix}
  } \right.
  =\left\{ {
\begin{matrix}
\begin{split}
   &{{F_{\tilde E_{\rm I}}}\left( \varepsilon_{j-i+\delta+1} \over \lambda \right)-{F_{\tilde E_{\rm I}}}\left( \varepsilon_{j-i+\delta} \over \lambda \right), \quad \text{if} \quad i- \delta \le j<L-\delta}\\
  &{1- {F_{\tilde E_{\rm I}}}\left( \varepsilon_{L-i} \over \lambda \right), \quad \text{if} \quad j=L-\delta}\\
  &{0, \quad \text{Otherwise}}\\
\end{split}
\end{matrix}
  } \right..
\end{split}
\end{equation}
\hrulefill
\vspace*{4pt}
\end{figure*}

Let $\mathbf{Z} = ({T_{i,j}})$ denote the $(L+1) \times (L+1)$ state transition matrix of the MC. By using similar methods in \cite{row-stac}, we can easily verify that the MC transition matrix $\mathbf{Z}$ derived from the above MC model is irreducible and row stochastic. Thus, there must exist a unique stationary distribution $\pmb{\pi}$ that satisfies the following equation 
\begin{equation}\label{solve}
\pmb {\pi} =\left( {{{\pi} _{0}},{{\pi} _{1}}, \cdots, {{\pi} _{L}}} \right)^{T} = \left({\mathbf{Z}}\right)^{T} \pmb {\pi},
\end{equation}
where ${{\pi} _{i}}$, $i \in \left\{ {0,1, \cdots ,L} \right\}
$, is the $i$-th component of $\pmb {\pi}$ representing the stationary distribution of the $i$-th energy level at relay. The battery stationary distribution of relay can be solved from (\ref{solve}) and expressed as 
\begin{equation}\label{app-energy-distri}
 \pmb {\pi} = {\left( { \left({\mathbf{Z}}\right)^{T} - \mathbf{I} + \mathbf{B} }\right)^{ - 1}}\mathbf{b},
\end{equation}
where ${\mathbf{B}_{i,j}}=1, \forall i, j$ and $\mathbf{b}={(1,1, \cdots ,1)^T}$.
\subsection{System Throughput Analysis}
In this paper, we select system throughput as the system performance measure. For the considered WP-TWRN, the system throughput is defined as the sum of the average throughput of the two sources. The system throughput of the proposed PS-EA scheme can be expressed as
\begin{equation}\label{sum-throughput1}
\begin{split}
\Psi &=\sigma_B \Pr \left\{ {\Upsilon  = {\Upsilon _{\rm II}},{\gamma _A} > v_B} \right\} + \sigma_A \Pr \left\{ {\Upsilon  = {\Upsilon _{\rm II}},{\gamma _B} > v_A} \right\}\\
&=\sigma_B \sum\limits_{i = \delta }^L {{\pi _i}\Pr \left\{ {{\gamma _A} > v_B} \right\} }+ \sigma_A \sum\limits_{i = \delta }^L {{\pi _i}\Pr \left\{ {{\gamma _B} > v_A} \right\}},
\end{split}
\end{equation}
where $\sigma_A$ and $\sigma_B$ are the transmission rates of source $A$ and $B$ respectively and $v_A=2^{2\sigma_A}-1$, $v_B=2^{2\sigma_B}-1$ are the outage thresholds evaluated from the channel capacity. Moreover, the last equality holds according to the operation principles that the relay operates in Mode II only when the accumulated energy is higher than energy level $\delta$. In the following, we evaluate the two probability terms in (\ref{sum-throughput1}). The output SNRs given in (\ref{SNRA}), (\ref{SNRB}) can be further summarized in the form of $\gamma_A, \gamma_B={{aHG} \over {bH + cG + 1}}$ and $a,b,c$ are different constants corresponding to the SNRs at different sources. The two probability terms in (\ref{sum-throughput1}) can be easily obtained by first evaluating the general term $\phi \left( {a,b,c,v} \right) = \Pr \left\{ {{{aHG} \over {bH + cG + 1}} > v} \right\}$.

In \cite{Raymond-two-way-relay}, the authors evaluated the similar term $\phi \left( {a,b,c,v} \right)$ where $H$, $G$ are exponential random variables. In this paper, we extend the analysis in \cite{Raymond-two-way-relay} to a more general and complicated case, in which $H$ and $G$ are gamma random variable. In particular, the CDF of $H$ and PDF of $G$ are given by ${F_H}\left( x \right) = 1 - \sum\limits_{i = 0}^{{m_A} - 1} {{{{{\left( {{{{m_A}} \over {{\Omega _A}}}x} \right)}^i}} \over {i!}}} \exp \left( { - {{{m_A}} \over {{\Omega _A}}}x} \right)$, ${f_G}\left( x \right) = {{{{\left( {{{{m_B}} \over {{\Omega _B}}}} \right)}^{{m_B}}}} \over {\Gamma \left( {{m_B}} \right)}}{x^{{m_B} - 1}}\exp \left( { - {{{m_B}} \over {{\Omega _B}}}x} \right)$ \cite[(Eq.2.21)]{fadingchannels}.

\begin{figure*}[!t]
\begin{equation}\label{finalresult}
\begin{split}
\phi \left( {a,b,c,v} \right) = & {{2{{\left( {{{{m_B}} \over {{\Omega _B}}}} \right)}^{{m_B}}}} \over {\Gamma \left( {{m_B}} \right)}}\exp \left[ { - {v \over a}\left( {{{{m_A}c} \over {{\Omega _A}}} + {{{m_B}b} \over {{\Omega _B}}}} \right)} \right] \sum\limits_{i = 0}^{{m_B} - 1}{\sum\limits_{j = 0}^{{m_A} - 1} {\sum\limits_{k = 0}^j {{{{\binom {m_B-1}{i} \binom {j} {k}b^{{m_B} - i - 1}}{c^{j - k}}{{\left( {{{{m_A}} \over {{\Omega _A}}}} \right)}^j}} \over {j!{a^{{m_B} + j - k}}}}} } } \\
&\times {v^{{m_B} + j - i - 1}}{\left( {{{bc} \over a}v + 1} \right)^k}{\left[ {{{{{{m_A}} \over {{\Omega _A}}}v\left( {bcv + a} \right)} \over {{{{m_B}} \over {{\Omega _B}}}}}} \right]^{{{i - k + 1} \over 2}}} {K_{i - k + 1}}\left( {2\sqrt {{{{m_A}} \over {{\Omega _A}}}{{{m_B}} \over {{\Omega _B}}}\left( {{{bc} \over {{a^2}}}{v^2} + {v \over a}} \right)} } \right).
\end{split}
\end{equation}
\hrulefill
\vspace*{4pt}
\end{figure*}

For the purpose of brevity, we omit the details of the tedious derivation. With the above CDF and PDF, by using similar methods in \cite{Raymond-two-way-relay} and the integral in \cite[(Eq.3.471.9)]{Tableofintegral}, the final result is given in (\ref{finalresult}) on top of the next page. We can now obtain a closed-from expression for the system throughput of the proposed PS-EA scheme given by
\begin{equation}\label{sum-throughput}
\begin{split}
\Psi=&\sigma_B \sum\limits_{i = \delta }^L {{\pi _i}\phi \left( {{{\overline \gamma  }_R}{{\overline \gamma  }_B},{{\overline \gamma  }_R} + {{\overline \gamma  }_A},{{\overline \gamma  }_B},v_B} \right) }\\
&+ \sigma_A \sum\limits_{i = \delta }^L {{\pi _i} \phi \left( {{{\overline \gamma  }_R}{{\overline \gamma  }_A},{{\overline \gamma  }_A},{{\overline \gamma  }_R} + {{\overline \gamma  }_B},v_A} \right)}.
\end{split}
\end{equation}

\begin{remark}
From the above analysis, we can see that increasing power splitting ratio $\lambda$ will lead to two conflicting impacts on the system performance. On one hand, it decreases the output SNRs and reduces the system throughput since less power is split for information forwarding. On the other hand, more energy is harvested during the operation when the power splitting ratio increases, consequently, the probability that relay opts Mode II is higher and the system throughput grows. We can thus deduce that there must exist an optimal value of $0<\lambda<1$ such that the system throughput is maximized.

The parameter $\delta$ also influences the system performance significantly. When the value of $\delta$ increases, the probability that relay works in Mode II is lower and the system performance is confined. On the other hand, $P_R = 2\delta\varepsilon _1$ increases as $\delta$ grows. With higher transmit power of the relay, the output SNRs of both sources are increased and the overall throughput is increased. Thus we can conclude that there should also exist an optimal $\delta = {P_R \over {2\varepsilon_1}} \in \left\{ 1, 2, \cdots, L \right\}$ such that the system throughput is maximized. Unfortunately, it is difficult for us to further jointly optimize $\lambda$ and $\delta$ due to the complex structure of the MC model and the complicated expression of system throughput. However, optimal solutions can be easily obtained via numerical exhaustive search with the derived analytical expression given in (\ref{sum-throughput}).
\end{remark}

\section{Numerical Results}
In this section, we present some numerical results to validate and illustrate the above theoretical analysis. In order to capture the effect of path-loss, we use the model ${\Omega _{XY}} = {1 \over {1 + {d_{XY}^\alpha}}}$, where ${\Omega _{XY}}$ is the average channel power gain between node $X$ and $Y$, ${d_{XY}}$ denotes the distance between node $X$ and $Y$, and $\alpha  \in \left[ {2,5} \right]$ is the path-loss factor. For simplicity, we consider a linear topology such that the sources and relay are located in a straight line. In all the following simulations, we let $P_A=P_B=P$ and $\sigma_A=\sigma_B=\sigma$. We set the distances ${d_{AB}} = 20$m and $d_{AR}=8$m, the path-loss factor $\alpha  = 2$, the severity parameters $m_A=m_B=2$, the battery capacity $C=0.2$, the noise powers ${N_0} = 2N_1=2N_2= - 80$dBm and the energy conversion efficiency $\eta  = 0.5$.

\begin{figure}
\centering \scalebox{0.5}{\includegraphics{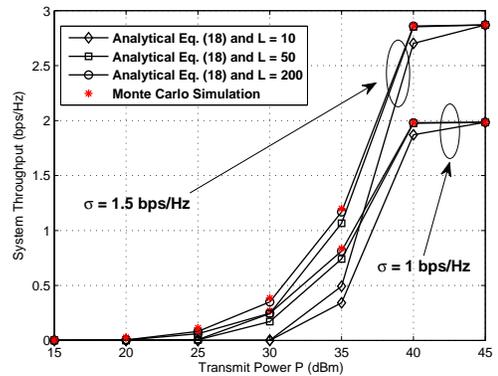}}
\caption{The system throughput of the proposed PS-EA scheme versus transmit power $P$ for different energy level $L$ and bit rate $\sigma$ where ${P_R \over 2} = 20 \% C$ and $\lambda=0.9$.\label{fig:montsimu1}}
\end{figure}

We first compare the analytical system throughput derived in (\ref{sum-throughput}) with the Monte Carlo simulation result. In order to explore the impact of $L$ in the MC model, we set the required forwarding energy of the relay to be 20\% of the total battery capacity which is independent of $L$. From Fig. \ref{fig:montsimu1}, we can see that the analytical results approach the simulation results as $L$ increases. Specifically, the analytical results for $L=200$ coincide well with the Monte Carlo simulation which validates our theoretical analysis presented in Sec. III. Furthermore, we can also observe that the accuracy of the MC model improves as the transmit power grows. This is understandable since for the same degree of precision, smaller intervals between adjacent energy levels (i.e. lager $L$ for a given $C$) are required to accurately capture the amount of charging energy when the harvested energy is low. At last, the system throughput becomes saturated when the transmit power is sufficiently large. This is due to the storage capacity constraint of the relay as well as the transmission rate bound. As the analytical results agree well with the simulation results and for the purpose of simplicity, in the following, we will only plot the analytical results of the proposed PS-EA scheme when $L=200$.

\begin{figure}
\centering \scalebox{0.5}{\includegraphics{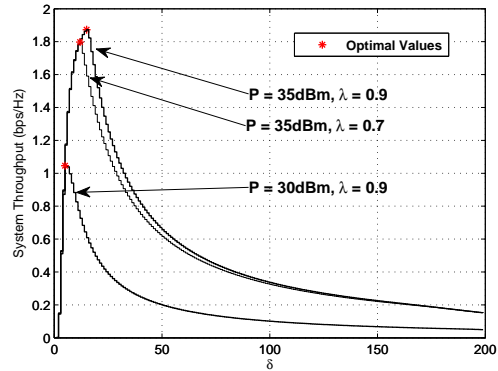}}
\caption{The system throughput of the proposed PS-EA scheme versus $\delta$ for different source transmit power $P$ and power splitting ratio $\lambda$.\label{fig:lplot}}
\end{figure}

In Fig. \ref{fig:lplot}, we depict the system throughput of the proposed PS-EA scheme versus $\delta$ for different source transmit power and power splitting ratio. Recall that $\delta  \in \left\{ 1, 2, \cdots, L \right\}$ is defined as the energy level of the relay transmit power $P_R$. Since the value of $\delta$ is discrete, we plot the throughput performance curves in stairs manner. We can observe that there exist an optimal value of $\delta$ that maximizes the system throughput for all the considered cases. It validates our deduction in Remark 1. Moreover, we can see that the optimal value of $\delta$ shifts to the right as the transmit power $P$ increases. This is because the relay can harvest more energy on average when the transmit power of the source increases and a higher transmit power of relay can be supported for a better system performance. We can also observe that the optimal value of $\delta$ decreases as the power splitting ratio $\lambda$ decreases. This is understandable since the relay harvests less energy as the power splitting ratio reduces and a smaller transmit power should be used.

\begin{figure}
\centering \scalebox{0.5}{\includegraphics{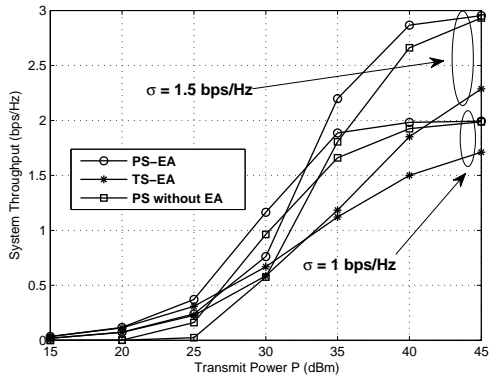}}
\caption{The system throughput of the proposed PS-EA scheme, its counterpart TS-EA scheme and the PS scheme without EA with their optimal settings.\label{fig:compare}}
\end{figure}

In Fig. \ref{fig:compare}, we compares the optimal throughput of the proposed PS-EA scheme, its counterpart time switching based energy accumulation (TS-EA) scheme and the conventional PS scheme without EA. The optimal throughput of the proposed PS-EA scheme can be obtained by jointly optimizing the parameters $0<\lambda<1$ and $\delta \in \left\{1,2,\cdots,L\right\}$. We perform a two-dimensional exhaustive search from the analytical expression to achieve the optimal performance. In the TS-EA scheme, the time switching scheme is implemented such that the relay can only perform either energy harvesting or information forwarding operations in each transmission block. Specifically, energy harvesting operation is the same as Mode I operation in the proposed PS-EA scheme, while in the information forwarding operation, all the received signal is delivered to the information receiver and nothing is used for energy harvesting. Thus, the optimal system throughput of TS-EA scheme can be attained by finding the optimal transmit power of the relay. For the conventional PS scheme without EA, the relay exhausts the harvested energy to forward information during each transmission block. The system throughput can thus be optimized by the optimal power splitting ratio. From Fig. 3, we can see that TS-EA scheme only outperforms the PS scheme without EA at low and medium SNRs range. Although the conventional PS does not accumulate the harvested energy, it is still superior to TS-EA scheme. This is mainly because the PS technique is more efficient than the TS technique since it enables the simultaneous energy harvesting and information forwarding. More importantly, the proposed PS-EA scheme can outperform the two benchmark schemes in all simulated cases.

\section{Conclusions}
In this paper, we proposed a power splitting-based energy accumulation (PS-EA) scheme for wireless-powered two-way relay networks (WP-TWRNs). We modeled the dynamic charging and discharging behaviors of the finite-capacity relay battery by a a finite-state Markov chain (MC). We proposed a novel operation mode-based approach to evaluate the transition probability and stationary distribution of the MC. We then derived the system throughput of the proposed PS-EA scheme, which is defined as the sum-throughput of two sources, over Nakagami-m fading channels. Numerical results validated the theoretical analysis and demonstrated the impact of various system parameters on the performance. Results showed that the proposed PS-EA scheme outperforms its counterpart time-switching based energy accumulation scheme and the PS scheme without energy accumulation in all the considered cases.

\ifCLASSOPTIONcaptionsoff
  \newpage
\fi


\bibliographystyle{IEEEtran}
\bibliography{References}

\begin{thebibliography}{10}
\providecommand{\url}[1]{#1}
\csname url@samestyle\endcsname
\providecommand{\newblock}{\relax}
\providecommand{\bibinfo}[2]{#2}
\providecommand{\BIBentrySTDinterwordspacing}{\spaceskip=0pt\relax}
\providecommand{\BIBentryALTinterwordstretchfactor}{4}
\providecommand{\BIBentryALTinterwordspacing}{\spaceskip=\fontdimen2\font plus
\BIBentryALTinterwordstretchfactor\fontdimen3\font minus
  \fontdimen4\font\relax}
\providecommand{\BIBforeignlanguage}[2]{{%
\expandafter\ifx\csname l@#1\endcsname\relax
\typeout{** WARNING: IEEEtran.bst: No hyphenation pattern has been}%
\typeout{** loaded for the language `#1'. Using the pattern for}%
\typeout{** the default language instead.}%
\else
\language=\csname l@#1\endcsname
\fi
#2}}
\providecommand{\BIBdecl}{\relax}
\BIBdecl

\bibitem{powersplitting3}
A.~A. Nasir, X.~Zhou, S.~Durrani, and R.~A. Kennedy, ``Relaying protocols for
  wireless energy harvesting and information processing,'' \emph{IEEE Trans. on
  Wireless Communications}, vol.~12, no.~7, pp. 3622--3636, 2013.

\bibitem{CHEN_TWC}
H.~Chen, Y.~Li, Y.~Jiang, Y.~Ma, and B.~Vucetic, ``Distributed power splitting
  for swipt in relay interference channels using game theory,'' \emph{IEEE
  Trans. Wireless Commun.}, vol.~14, no.~1, pp. 410--420, Jan 2015.

\bibitem{Raymond-two-way-relay}
R.~H.~Y. Louie, Y.~Li, and B.~Vucetic, ``Practical physical layer network
  coding for two-way relay channels: performance analysis and comparison,''
  \emph{IEEE Trans. on Wireless Communications}, vol.~9, no.~2, pp. 764--777,
  2010.

\bibitem{TW-TS-1}
Y.~Liu, L.~Wang, M.~Elkashlan, T.~Q. Duong, and A.~Nallanathan, ``Two-way
  relaying networks with wireless power transfer: Policies design and
  throughput analysis,'' in \emph{2014 IEEE Global Communications Conference
  (GLOBECOM)}, 2014, pp. 4030--4035.

\bibitem{TW-TS-2}
K.~Xiong, P.~Fan, and K.~B. Letaief, ``Time-switching based swpit for
  network-coded two-way relay transmission with data rate fairness,'' in
  \emph{2015 IEEE International Conference on Acoustics, Speech and Signal
  Processing (ICASSP)}, 2015, pp. 5535--5539.

\bibitem{TW-TS-3}
C.~Huang, P.~Sadeghi, and A.~A. Nasir, ``B{E}{R} performance analysis and
  optimization for energy harvesting two-way relay networks,'' in \emph{2016
  Australian Communications Theory Workshop (AusCTW)}, 2016, pp. 65--70.

\bibitem{TW-PS-1}
Z.~Chen, B.~Xia, and H.~Liu, ``Wireless information and power transfer in
  two-way amplify-and-forward relaying channels,'' in \emph{2014 IEEE Global
  Conference on Signal and Information Processing (GlobalSIP)}, 2014, pp.
  168--172.

\bibitem{TW-PS-2}
X.~Lu, W.~Xu, S.~Li, Z.~Liu, and J.~Lin, ``Simultaneous wireless information
  and power transfer for cognitive two-way relaying networks,'' in \emph{2014
  IEEE 25th Annual International Symposium on Personal, Indoor, and Mobile
  Radio Communication (PIMRC)}, 2014, pp. 748--752.

\bibitem{TW-PS-3}
Z.~Wang, Z.~Chen, Y.~Yao, B.~Xia, and H.~Liu, ``Wireless energy harvesting and
  information transfer in cognitive two-way relay networks,'' in \emph{2014
  IEEE Global Communications Conference (GLOBECOM)}, 2014, pp. 3465--3470.

\bibitem{TW-PS-4}
J.~Men, J.~Ge, C.~Zhang, and J.~Li, ``Joint optimal power allocation and relay
  selection scheme in energy harvesting asymmetric two-way relaying system,''
  \emph{IET Communications}, vol.~9, no.~11, pp. 1421--1426, 2015.

\bibitem{row-stac}
I.~Krikidis, S.~Timotheou, and S.~Sasaki, ``R{F} energy transfer for
  cooperative networks: Data relaying or energy harvesting?'' \emph{IEEE
  Communications Letters}, vol.~16, no.~11, pp. 1772--1775, November 2012.

\bibitem{TS-accumulation}
Y.~Gu, H.~Chen, Y.~Li, and B.~Vucetic, ``A discrete time-switching protocol for
  wireless-powered communications with energy accumulation,'' in \emph{2015
  IEEE Global Communications Conference (GLOBECOM)}, 2015.

\bibitem{five-multi-relay}
I.~Krikidis, ``Relay selection in wireless powered cooperative networks with
  energy storage,'' \emph{IEEE Journal on Selected Areas in Communications},
  vol.~33, no.~12, pp. 2596--2610, 2015.

\bibitem{YIFAN11}
Y.~Gu, H.~Chen, Y.~Li, and B.~Vucetic, ``Distributed multi-relay selection in
  wireless-powered cooperative networks with energy accumulation,'' in
  \emph{2016 IEEE International Conference on Communications (ICC)}, 2016.

\bibitem{Tableofintegral}
A.~Jeffrey and D.~Zwillinger, \emph{Table of Integrals, Series, and Products},
  ser. Table of Integrals, Series, and Products Series.\hskip 1em plus 0.5em
  minus 0.4em\relax Elsevier Science, 2007.

\bibitem{Zhong_Tcom_2015_Wireless}
C.~Zhong, X.~Chen, Z.~Zhang, and G.~Karagiannidis, ``Wireless powered
  communications: Performance analysis and optimization,'' \emph{To appear in
  IEEE Trans. Commun.}, 2015.

\bibitem{Huang_TWC_2008_Wire}
W.-J. Huang, Y.-W. Hong, and C.-C. Kuo, ``Lifetime maximization for
  amplify-and-forward cooperative networks,'' \emph{IEEE Trans. Wireless
  Commun.}, vol.~7, no.~5, pp. 1800--1805, May 2008.

\bibitem{fadingchannels}
M.~Simon and M.~Alouini, \emph{Digital Communication over Fading Channels},
  ser. Wiley Series in Telecommunications and Signal Processing.\hskip 1em plus
  0.5em minus 0.4em\relax Wiley, 2005.

\bibitem{sum-two-gamma}
G.~K. Karagiannidis, N.~C. Sagias, and T.~A. Tsiftsis, ``Closed-form statistics
  for the sum of squared nakagami-m variates and its applications,'' \emph{IEEE
  Trans. on Communications}, vol.~54, no.~8, pp. 1353--1359, 2006.

\end{thebibliography}

%

\end{document}